\documentclass[pdftex,twocolumn,epjc3]{svjour3}          

\RequirePackage[T1]{fontenc}

\smartqed  

\RequirePackage{graphicx}
\RequirePackage{mathptmx}      
\RequirePackage{flushend}
\RequirePackage[numbers,sort&compress]{natbib}
\RequirePackage[colorlinks,citecolor=blue,urlcolor=blue,linkcolor=blue]{hyperref}
\RequirePackage{xcolor}
\RequirePackage[normalem]{ulem}
\RequirePackage{amsmath}

\newcommand{\etal}{\textit{et}\space\textit{al}.\space}

\journalname{Eur. Phys. J. A}

\begin{document}

\title{A look at multiparticle production via modified combinants}

\author{Han Wei Ang\thanksref{e1,addr1}
        \and
        Aik Hui Chan\thanksref{e2,addr1}
        \and
        Mahumm Ghaffar\thanksref{e3,addr2}
        \and
        Maciej Rybczy\'nski\thanksref{e4,addr3}
        \and
        Grzegorz Wilk\thanksref{e5,addr4}
        \and
        Zbigniew W\l odarczyk\thanksref{e6,addr3}
}

\thankstext{e1}{e-mail: ang.h.w@u.nus.edu}
\thankstext{e2}{e-mail: phycahp@nus.edu.sg}
\thankstext{e3}{e-mail: mghaffar@mun.ca}
\thankstext{e4}{e-mail: maciej.rybczynski@ujk.edu.pl}
\thankstext{e5}{e-mail: grzegorz.wilk@ncbj.gov.pl}
\thankstext{e6}{e-mail: zbigniew.wlodarczyk@ujk.edu.pl}

\institute{Department of Physics, National University of Singapore, Singapore 17551\label{addr1}
          \and
          Department of Physics and Physical Oceanography, Memorial University of Newfoundland, St. John's, NL A1B 3X7, Canada\label{addr2}
          \and
          Institute of Physics, Jan Kochanowski University, 25-406 Kielce, Poland\label{addr3}
          \and
          National Centre for Nuclear Research, Department of Fundamental Research, 02-093 Warsaw, Poland\label{addr4}
}

\date{Received: date / Accepted: date}

\maketitle

\begin{abstract}
As shown recently, one can obtain additional information from the measured charged particle multiplicity distributions, $P(N)$, by investigating the so-called modified combinants, $C_j$, extracted from them. This information is encoded in the observed specific oscillatory behaviour of $C_j$, which phenomenologically can be described only by some combinations of compound distributions based on the Binomial Distribution. So far this idea has been checked in $pp$ and $e^+e^-$ processes (where observed oscillations are spectacularly strong). In this paper we continue observation of multiparticle production from the modified combinants perspective by investigating dependencies of the observed oscillatory patterns on type of colliding particles, their energies and the phase space where they are observed.
We also offer some tentative explanation based on different types of compound distributions and stochastic branching processes.
\end{abstract}

\section{Introduction}
\label{sect.intro}
Multiplicity distributions (MDs) of high energy collisions have been extensively studied in the field of multiparticle production.  It is one of the first observables to be determined in new high-energy experiments. This is partly due to the ease with which such information can be obtained, and also because MDs contain useful information on the underlying production processes. Due to the inability of perturbative Quantum Chromodynamics (pQCD) to provide a complete theoretical account for the observed MDs incorporating both the hard and soft processes, various phenomenological approaches had to be adopted. These can range from dynamical approaches in the form of coloured string interactions \cite{Lund} and dual-parton model \cite{DualParton}, to geometrical approaches  \cite{GBM1,GBM2} resulting in the fireball model \cite{Fireball}, stochastic approaches \cite{GMD, Chewetal, p-pbar} modelling high energy collision as branching \cite{GMD, Chewetal, p-pbar} or clans \cite{ClanModel}.

The myriad of stochastic models since proposed have described the experimental data well with very reasonable $\chi^2/dof$ values. Amongst the numerous proposed distributions, the Negative Binomial Distribution (NBD) and its variants are the most ubiquitous \cite{NBD}. However, as has been proposed recently \cite{JPG, MWW1, RWW-Odessa, 3componentNBD}, a good fit to the MD from a statistical distribution is only one aspect of a full description of the multi-faceted set of information derivable from the MDs. A more stringent requirement before any phenomenological model is considered viable is to also reproduce the oscillatory behaviour seen in the so called modified combinant, $C_j$, which can be derived from experimental data. In fact, this phenomenon is observed not only in $pp$ collisions discussed in \cite{JPG, MWW1, RWW-Odessa, 3componentNBD} but also, as demonstrated recently in \cite{e+e-}, in $e^+e^-$ annihilation processes. Such oscillations may be therefore indicative of additional information on the multiparticle production process, so far undisclosed. Specifically, the periodicity of the oscillations of modified combinants derived from experimental data is suggestive.

It is in this spirit that this study sets forth to understand the effects of the collision systems and various experimental observables on the period and extent of oscillations in $C_j$. In Section \ref{sect.mod.combinant}, the concept of \textit{modified combinant} will be reviewed in light of its connection to the earlier concept of \textit{combinant} \cite{ST,Combinants,Combinants2}. From this link, an attempt is made on the potential interpretation of modified combinant applied in the context of multi-particle production.
Section \ref{sect.dependence} discusses the problem of dependence on collision system whereas Section \ref{sect.oscillation.dependence} discusses the effect various experimental variables have on the modified combinant oscillations and summarises the key points observed.

Our concluding remarks are contained in Section \ref{sect.conclusion} together with a tentative proposal of employing the characteristics of oscillations in experimental modified combinants to distinguish between different collision types. Some explanatory material is presented in appendices: \ref{appendixA} presents the relationship between $C_j$ and the $K_q$ and $F_q$ moments that are more familiar to the particle physics community whereas \ref{appendixB} shows the possible origin of the observed oscillations of $C_j$ based on the stochastic approach to the particle production processes.

\section{Modified Combinant and Combinant}
\label{sect.mod.combinant}

Statistical distributions describing charged particle multiplicity are normally expressed in terms of their generating function, $G(z) = \sum_{N=0}^{\infty} P(N)z^N$, or in terms of their probability function $P(N)$. One other way to characterise a statistical distribution is a recurrent form involving only adjacent values of $P(N)$ for the production of $N$ and $(N+1)$ particles,
\begin{equation}
    (N+1)P(N+1) = g(N)P(N).
\label{recurrence.P(N)}
\end{equation}
Cast in this form, every $P(N)$ value is assumed to be determined only by the next lower $P(N-1)$ value. In other words, the link to other $P(N-j)$'s for $j>1$ is indirect. In addition, the eventual algebraic form of the $P(N)$ is determined by the function $g(N)$. In its simplest form, one can assume $g(N)$ to be linear in $N$, such that
\begin{equation}
    g(N) = \alpha + \beta N  \label{g(n)}.
\end{equation}
Some prominent distributions have been defined in this form. For example, when $ \beta = 0$ one gets the Poisson Distribution (PD). The Binomial Distribution (BD) arises for $ \beta < 0$ and $ \beta > 0$ results in the Negative Binomial Distribution (NBD). While conceptualising a phenomenological model, the form of $g(N)$ can be modified accordingly to describe the experimental data, cf., for example,  \cite{Hoang1987, Zborovsky2011}.

\begin{table}[h]
\caption {Distributions $P(n)$ used in this work: Poisson (PD), Negative Binomial (NBD) and Binomial (BD),  their generating functions $G(z)$ and modified combinants $C_j$ emerging from them.} \vspace*{0.2cm}
\begin{center}
\begin{tabular}{|l|c|c|c|}
\hline
&                                                                     &           & \\
          &~ $P(N)$ ~ &~ $G(z)$ ~ & ~$C_j$~ \\
&                                                                     &           & \\
\hline
&                                                                     &           & \\
PD   &  $\frac{\lambda^N}{N!} \exp( - \lambda)$                       &  $\exp[\lambda (z - 1)]$                   & $\delta_{j0}$                       \\
&                                                                     &           & \\
\hline
&                                                                     &           & \\
NBD  &  ${N+k-1\choose N} p^N (1 - p)^k$                                  &  $\left( \frac{1 - p}{1 - pz}\right)^k$    & $\frac{k}{\langle N\rangle} p^{j+1}$ \\
&                                                                     &           & \\
\hline
&                                                                     &           & \\
BD   &  ${ K\choose N} p^N (1 - p)^{K-N}$                             &  $(pz + 1 - p)^K$                          & $\frac{-K}{\langle N\rangle} \left( \frac{p}{p - 1}\right)^{j+1}$\\
&                                                                     &           & \\
\hline
\end{tabular}
\end{center}
\label{Table_1}
\end{table}

However, the direct dependence of $P(N+1)$ on only $P(N)$, as seen in Eq. \ref{recurrence.P(N)}, seems unnecessarily restrictive. This constraint can be further relaxed, by writing the probability function connecting all smaller values of $P(N-j)$ as follows \cite{ST},
\begin{equation}
    (N + 1)P(N + 1) = \langle N\rangle \sum^{N}_{j=0} C_j P(N - j).     \label{eqn.Cj.recur}
\end{equation}
The coefficients $C_j$ are known as the \textit{modified combinants} and forms the core of this study. They are related to the combinants $C^*_j$ first defined for the study of boson production models \cite{Combinants,Combinants2} by the following relation \cite{JPG}:
\begin{equation}
    C_j = \frac{(j+1)}{\langle N\rangle}C^*_{j+1}.  \label{relation.to.combinant}
\end{equation}
Combinants were first introduced to quantify the extent any distributions deviate from a Poisson distribution. For the Poisson distribution $C_0 = 1$ and $C_{j>1} = 0$. In this way, any non-zero $C_j$ at higher orders indicate a deviation from the Poisson distribution.

From Eq. (\ref{eqn.Cj.recur}), two obvious interpretations for $C_j$ follow. First, there is a one-to-one map between $C^*_j$ to $C_j$ via Eq. (\ref{relation.to.combinant}). Modified combinants can be interpreted as a proxy to the extent of deviation from a Poisson distribution at different higher orders.
Secondly, $C_j$'s are the normalized weights in the series for the value of $(N+1)P(N+1)$. This can be interpreted as the "memory" which $P(N+1)$ has of the $P(N-j)$ term. In other words, the modified combinants are the weights in which all earlier $P(N-j)$ values has on the current probability. In this interpretation the links between $P(N+1)$ to all $P(N-j)$ values are clearly established.

One further notes that since $C_j$'s are expressed in terms of the probability function in Eq. (\ref{eqn.Cj.recur}), it may be reasonable to attempt casting the modified combinant in terms of the generating function $G(z) = \sum_N P(N)z^N$. Such an expression is immensely useful should a theoretical distribution avail itself to describe experimental data. In this case, $C_j$ can be expressed as follows:
\begin{equation}
    \langle N \rangle C_j  = \frac{1}{j!}\frac{d^{j+1}\ln G(z)}{dz^{j+1}}\Bigg |_{z=0}.
\label{Cj.in.Gz}
\end{equation}
Modified combinants for some prominent distributions are shown in Tab. \ref{Table_1}.
Note that the generating functions of NBD and BD are in fact some quasi-power functions of $z$ and as such can be written in the form of the corresponding Tsallis distribution \cite{WW-APPB},
\begin{equation}
G(z) = \exp_q[\langle N\rangle (1-z)] = [ 1 + (q-1)\langle N\rangle (1-z)]^{\frac{1}{1-q}} \label{qGz}
\end{equation}
where
\begin{eqnarray}
q-1 &=& \frac{1}{k} = \frac{p}{(1-p)\langle N\rangle}\qquad {\rm for~NBD},\label{qNBD}\\
q-1 &=& -\frac{1}{K} = - \frac{p}{\langle N\rangle}\qquad \quad {\rm for~BD}, \label{qBD}
\end{eqnarray}
whereas for $q\to 1$ in both cases we obtain $G(z)$ for PD. Eqs. (\ref{qNBD}) and (\ref{qBD}) allow to write  $C_j$ for all three distributions differentiated by the above choice of the paramater $q$ in one formula,
\begin{equation}
C_j = \frac{1}{(q-1)\langle N\rangle + 1}\left[ \frac{(q-1)\langle N\rangle}{(q-1)\langle N\rangle + 1}\right]^j. \label{qCj}
\end{equation}
Note that while for the PD and NBD coefficients $C_j$ are monotonic and positive functions of rank $j$, they strongly oscillate for the BD. This feature will be very important in all our further analysis here.

To understand the effects of various experimental variables on oscillations of modified combinants, a mathematical expression is required for calculating the value of $C_j$ given $P(N)$. From Eq. (\ref{eqn.Cj.recur}), it follows that
\begin{equation}
\langle N \rangle C_j = (j+1)\left[ \frac{P(j+1)}{P(0)} \right] - \langle N\rangle \sum^{j-1}_{i=0}C_i \left[ \frac{P(j-i)}{P(0)} \right].
\label{rCj}
\end{equation}
Note that Eq. (\ref{rCj}) will require $P(0) > 0$ which is often the case as most experimental data on non-single diffraction collision exhibits enhanced void probability \cite{MWW1, voidprob}. In the event that the void probability is not made available, it will be inferred from the normalization of probability.
\begin{figure}[t]
\begin{center}
\includegraphics[scale=0.7]{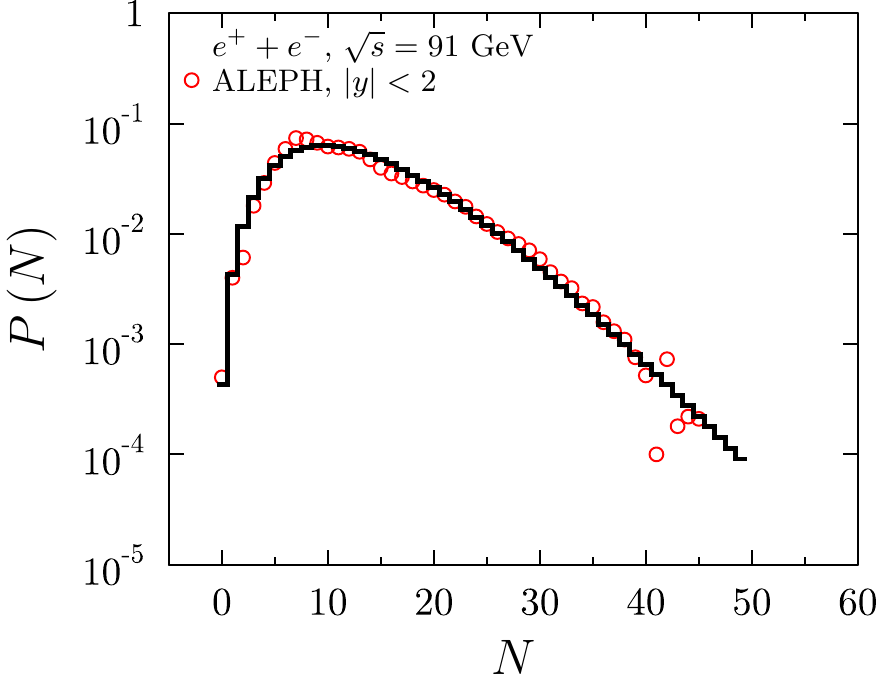}
\includegraphics[scale=0.7]{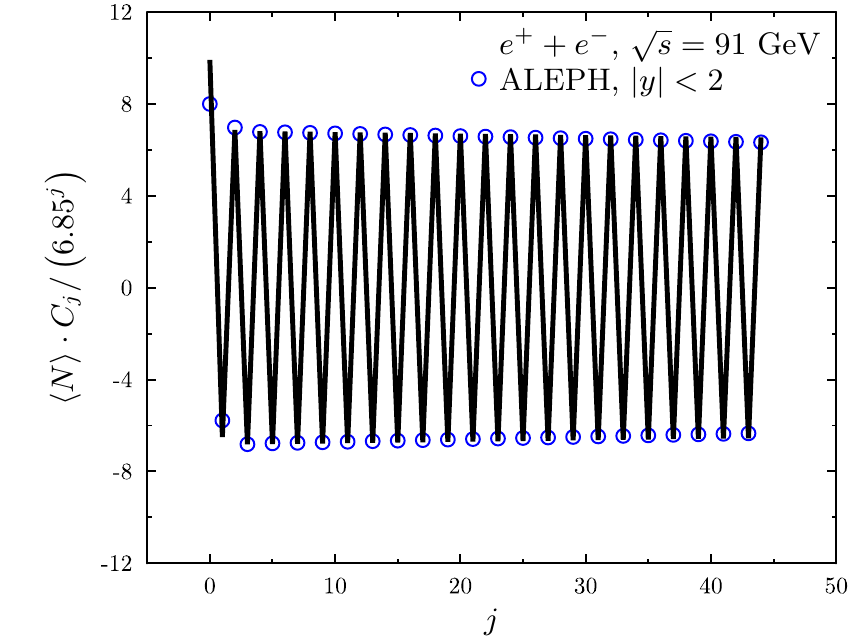}
\end{center}
\vspace{-5mm}
\caption{Top panel: Data on $P(N)$ measured in $e^+e^-$ collisions by the ALEPH experiment at $91$ GeV \cite{ALEPH} are fitted by the distribution obtained from the generating function given by Eq. (\ref{GBDNBD}) with parameters: $K=1$ and $p'=0.8725$ for the BD and $k=4.2$ and $p=0.75$ for the NBD.
 Bottom panel: the modified combinants $C_j$ deduced from these data on $P(N)$. They can be fitted by $C_j$ obtained from the same generating function  with the same parameters as used for fitting $P(N)$.}
\label{F1a}
\end{figure}
\begin{figure}[t]
\begin{center}
\includegraphics[scale=0.7]{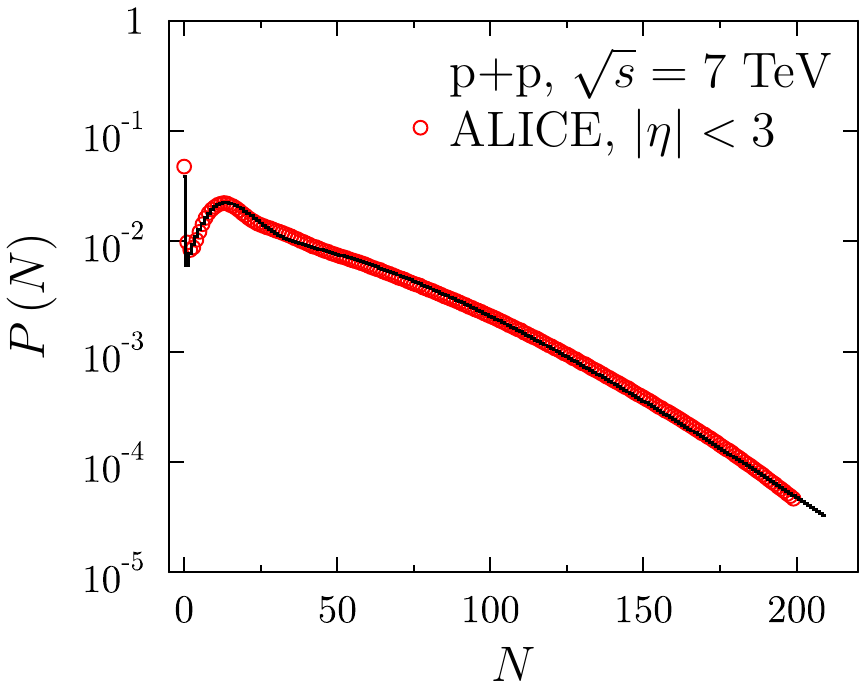}
\includegraphics[scale=0.7]{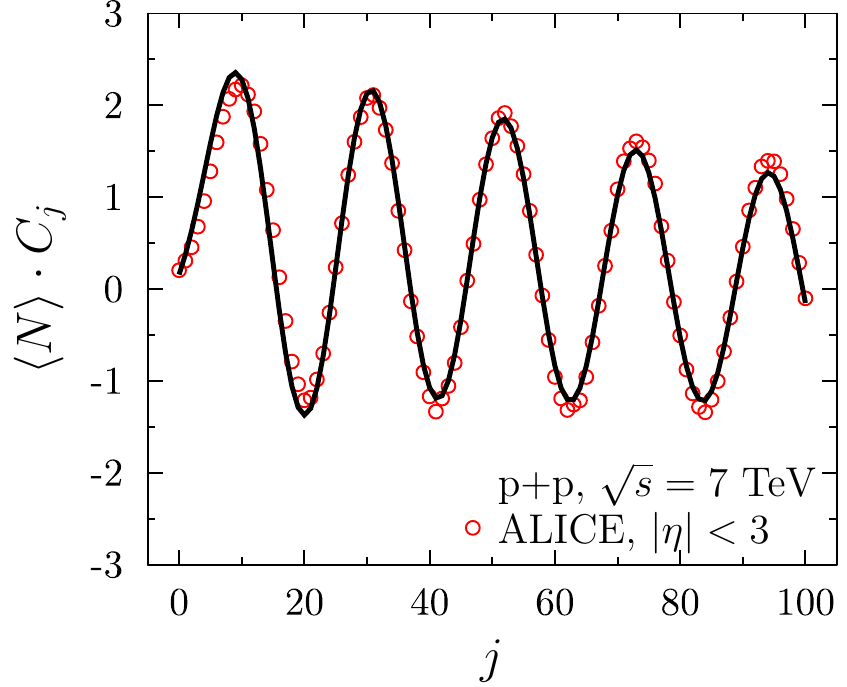}
\end{center}
\vspace{-5mm}
\caption{Top panel: Multiplicity distributions $P(N)$ measured in $pp$ collisions by ALICE \cite{ALICE}. Bottom panel: The corresponding modified combinants $C_j$. Data are fitted using sum of two compound distributions (BD+NBD) given by Eqs. (\ref{2-CBD}) and (\ref{2CBD}) with parameters:
 $K_1 = K_2 = 3$, $p_1 = 0.9$, $p_2 = 0.645$, $k_1 = 2.8$, $k_2 = 1.34$, $m_1 = 5.75$, $m_2 = 23.5$ , $w_1 = 0.24 $ and $w_2 = 0.76$ . }
\label{F1b}
\end{figure}

\section{Dependence of $C_j$ oscillations on collision systems}
\label{sect.dependence}

We shall start with a reminder of two distinct observed patterns of modified combinants, one observed
in $e^+e^-$ annihilation \cite{RWW-Odessa, e+e-} (cf. Fig. \ref{F1a}) and another observed in $pp$ scattering \cite{MWW1,RWW-Odessa} (cf. Fig. \ref{F1b}). In the first case we use the additivity property of modified combinants, i.e. that for a random variable composed of independent random variables, with its generating function given by the product of their generating functions, $G(x)=\prod_jG_j(x)$, the corresponding modified combinants are given by the sum of the independent components. For the $e^+e^-$ data we shall use then the generating function $G(z)$ of the multiplicity distribution $P(N)$ in which $N$ consists of both the particles from the BD ($N_{BD}$) and from the NBD ($N_{NBD}$):
\begin{equation}
N = N_{BD} + N_{NBD}. \label{NN}
\end{equation}
In this case generating function is
\begin{equation}
G(z)=G_{BD}(z)G_{NBD}(z) \label{GBDNBD}
\end{equation}
and multiplicity distribution can be written as
\begin{equation}
P(N) = \sum_{i=0}^{min\left\{ N,k\right\}} P_{BD}(i)P_{NBD}(N-i), \label{PbdPnbd}
\end{equation}
and the respective modified combinants are
\begin{equation}
\langle N\rangle C_j = \left< N_{BD}\right>C_j^{(BD)} + \left< N_{NBD}\right> C_j^{(NBD)}. \label{Cjbdnbd}
\end{equation}

Fig. \ref{F1a} shows the results of fits to both the experimentally measured \cite{ALEPH} multiplicity distributions and the corresponding modified combinants $C_j$ calculated from these data (cf. \cite{e+e-} for details).

\begin{figure*}[t]
\begin{center}
\includegraphics[scale=0.7]{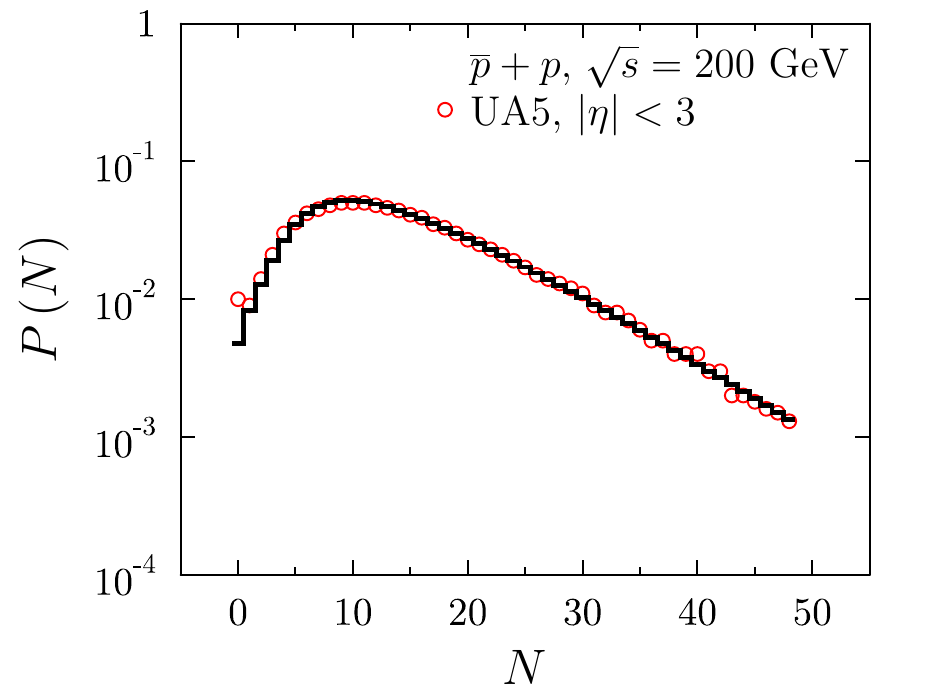}
\hspace{10mm}
\includegraphics[scale=0.7]{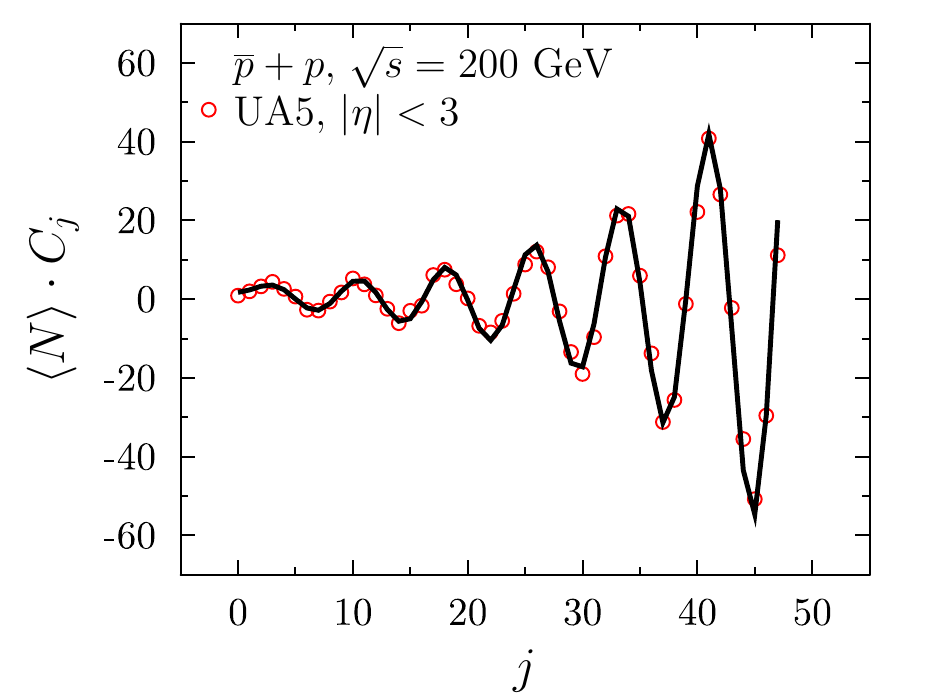}\\
\includegraphics[scale=0.7]{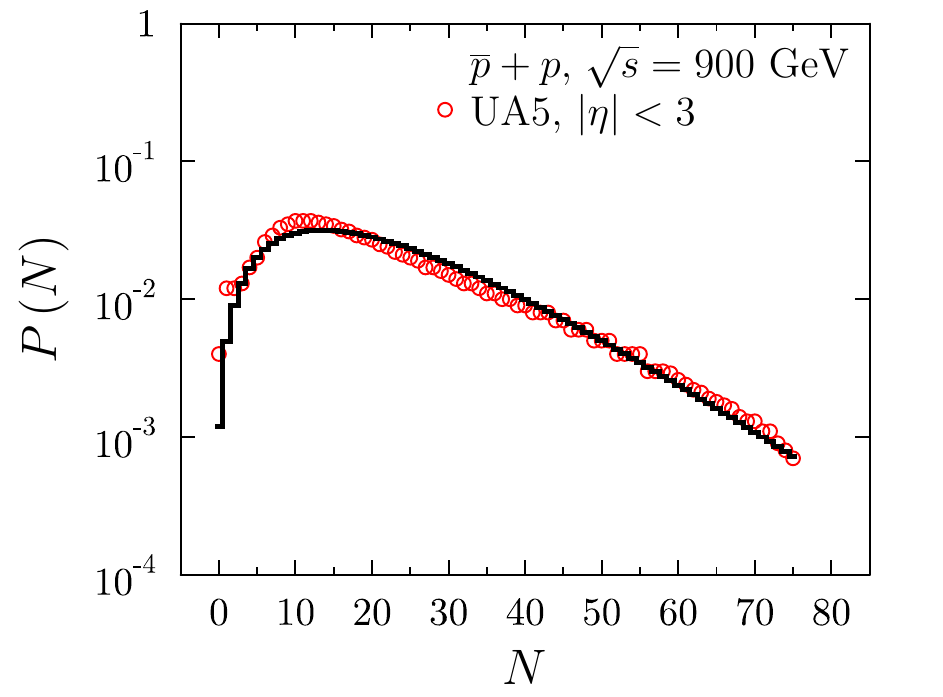}
\hspace{10mm}
\includegraphics[scale=0.7]{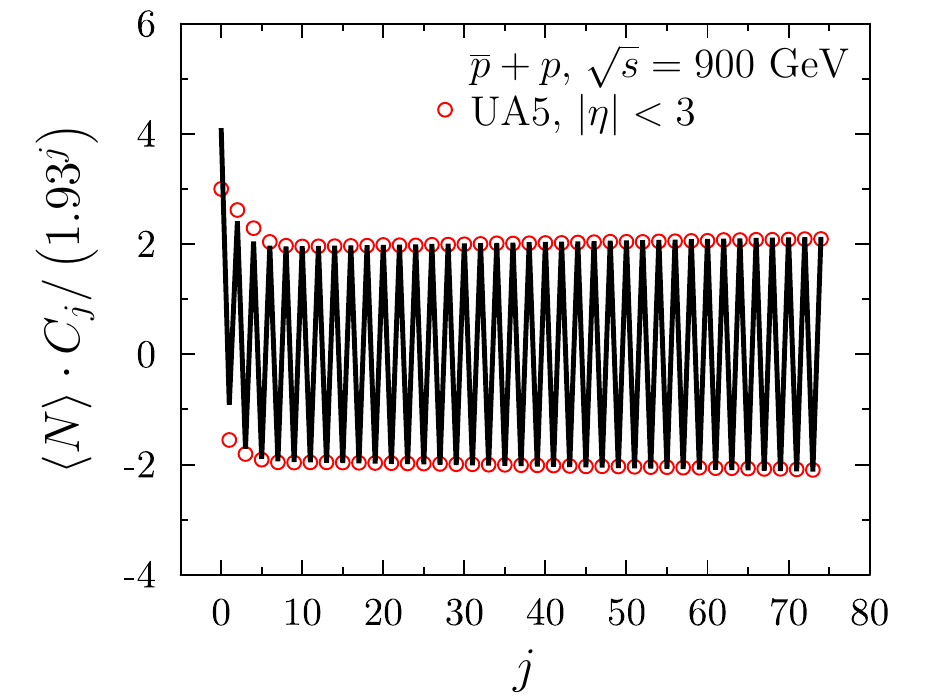}
\end{center}
\vspace{-5mm}
\caption{Left panels:  Multiplicity distributions $P(N)$ measured in $p \bar{p}$ collisions by UA5 experiment \cite{UA5900GeV}. Right panels:  The corresponding modified combinants $C_j$. Data at $900$ GeV are fitted  by the distribution obtained from the generating function given by the product $G(z) = G_{BD}(z)G_{NBD}(z)$ with parameters: $K = 1$ and $p' = 0.659$ for  the  BD and $k = 2.4$  and $p = 0.905$  for  the  NBD.  Data at $200$ GeV are fitted by the distribution obtained from the generating function given by $G(z)=G_{CD}(z)G_{NBD}(z)$ with parameters $K = 1$, $p' = 0.845$ and $\lambda = 4.6$ for the compound distribution CD (Binomial Distribution compound with Poisson, $BD\& PD$) and $k= 1.7$, $p = 0.875$ for the NBD. }
\label{F1c}
\end{figure*}

In the case of $pp$ collision the satisfactory agreement in fitting observed oscillatory pattern is obtained by using the sum of two Compound Binomial Distributions of BD with NBD type,
\begin{equation}
P(N) = \sum_{i=1,2} w_i h\left(N; p_i, K_i, k_i, m_i\right); \hspace{1cm} \sum_{i=1,2} w_i = 1
\label{2CBD}
\end{equation}
with the generating function of each component equal to
\begin{equation}
H(z) = \left[ p\left( \frac{1 - p'}{1 - p'z}\right)^k + 1 - p\right]^K; \hspace{1cm} p'=\frac{m}{m+k}
\label{2-CBD}
\end{equation}
As seen in Fig. \ref{F1b}, one gains satisfactory control over both the periods and amplitudes of the oscillations, as well as their behavior as a function of the rank $j$. More importantly one can reproduce the enhancement of void probability of $P(0)>P(1)$ in addition to fitting both the $P(N)$ and $C_j$.

The results presented in Figs. \ref{F1a} and \ref{F1b} suggest the possibility that the enhanced oscillatory behavior is, perhaps, a trait of the annihilation type of the process considered. To check this we turned to $p\bar{p}$ processes measured by UA5 \cite{UA5900GeV}. Fig. \ref{F1c} demonstrates that the outcome is rather intriguing and brings in new questions. At $900$ GeV one observes oscillatory pattern which follows that observed in annihilation process $e^+e^-$, and which can be fitted by the same kind of $P(N)$. However, the observed oscillatory pattern changes dramatically at $200$ GeV and resembles that seen before in the $pp$ collisions. It can still be fitted using generating function $G(z)$ given by Eq. (\ref{GBDNBD}) but with Binomial Distribution replaced by compound distribution CBD of the Binomial Distribution with Poisson distribution, i.e., by
\begin{equation}
G(z) = G_{CBD}(z)G_{NBD}(z) \label{CBD-NBD}
\end{equation}
where generating function for Compound Binomial Distribution (CBD) is given by
\begin{equation}
G_{CBD}(z) = \left[ p \exp[\lambda (z-1)] + 1 - p\right]^K. \label{G_CBD}
\end{equation}
Such replacement allows to preserve oscillating power of BD but, at the same time, to gain better control over the period of oscillations which is detemined by the mean multiplicity $\lambda$ in the PD \cite{MWW1}.

Note that the BD used at $900$ GeV can be considered as such compound distribution but with the PD replaced by $\delta_{N,1}$. It means therefore that, in order to fit the annihilation data at lower energies, one has to somehow smear out this delta-like behavior. In fact, one could as well use instead of the PD a NBD with large $k$ and $p$ such that $\lambda = kp/(1-p)$.

We close this Section by noting that the use of $G(z)$ in the form of Eq. (\ref{GBDNBD}) corresponds to a QCD-based approach based on stochastic branching processes used in \cite{e+e-}, the so-called Generalized Multiplicity distribution (GMD), with initial number of gluons given by a BD. The links between adopting a stochastic branching approach in the study of QCD phenomena has its roots in \cite{QCD.jets.MC}. In fact, similar approach was also formulated on general grounds in \cite{SBW} where it was shown that the stochastic birth process with immigration and with initial conditions given by BD results in the so-called Modified Negative Binomial Distribution (MNBD) (both approaches are presented in more detail in \ref{appendixB}). With more general choice of initial conditions, i.e., by replacing BD by some compound distribution  CD based on BD, one can, as presented here, describe also $p\bar{p}$ processes. However, in the case of $pp$ collisions this CD is more complicated (we have now $K = 3$ in our BD, which could, perhaps, correspond to $3$ valence quarks; additionally, to describe $P(N)$ we need in this case at least two such components).

\section{Dependence of $C_j$ oscillations on phase space being tested}
\label{sect.oscillation.dependence}

In addition to dependence on the collision system discussed above there are data \cite{CMS7TeV, ATLAS7TeV,ATLAS13TeV,ALICE8TeV.wide.eta,ALICE,UA5900GeV} (see also \cite{3componentNBD}) which allows to investigate the possible oscillatory behavior of $C_j$ in different pseudorapidity windows $|\eta|$, for different transverse momentum cuts $p_T$ and for different collision energies $\sqrt{s}$. We shall study them in this section.

\begin{figure}[t]
\begin{center}
\includegraphics[scale=0.75]{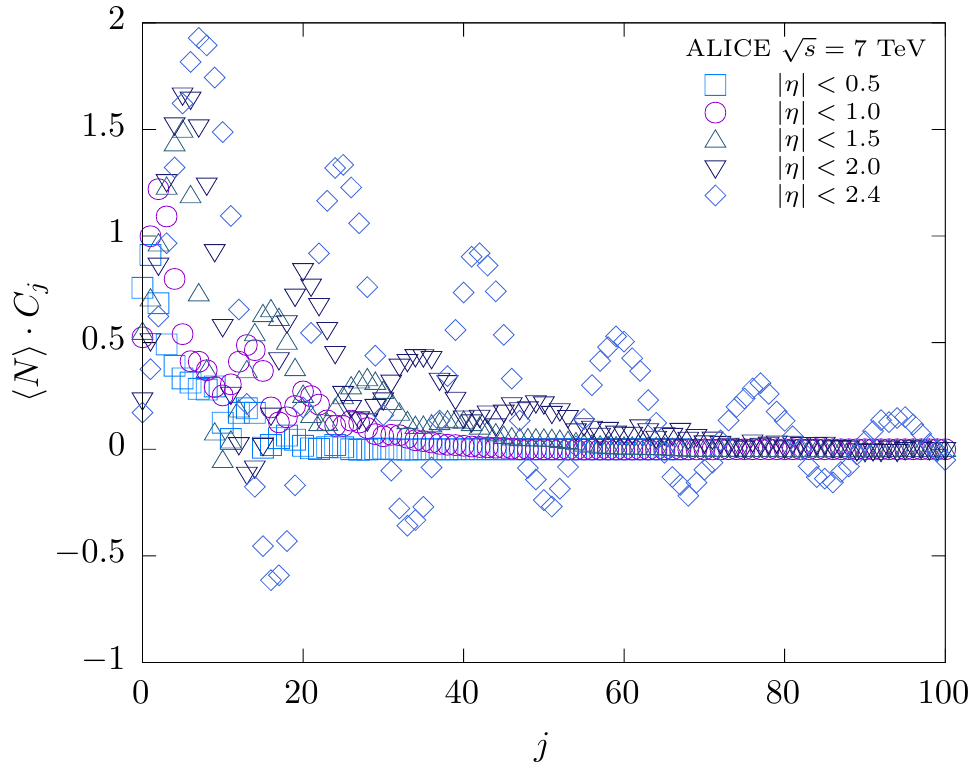}
\includegraphics[scale=0.75]{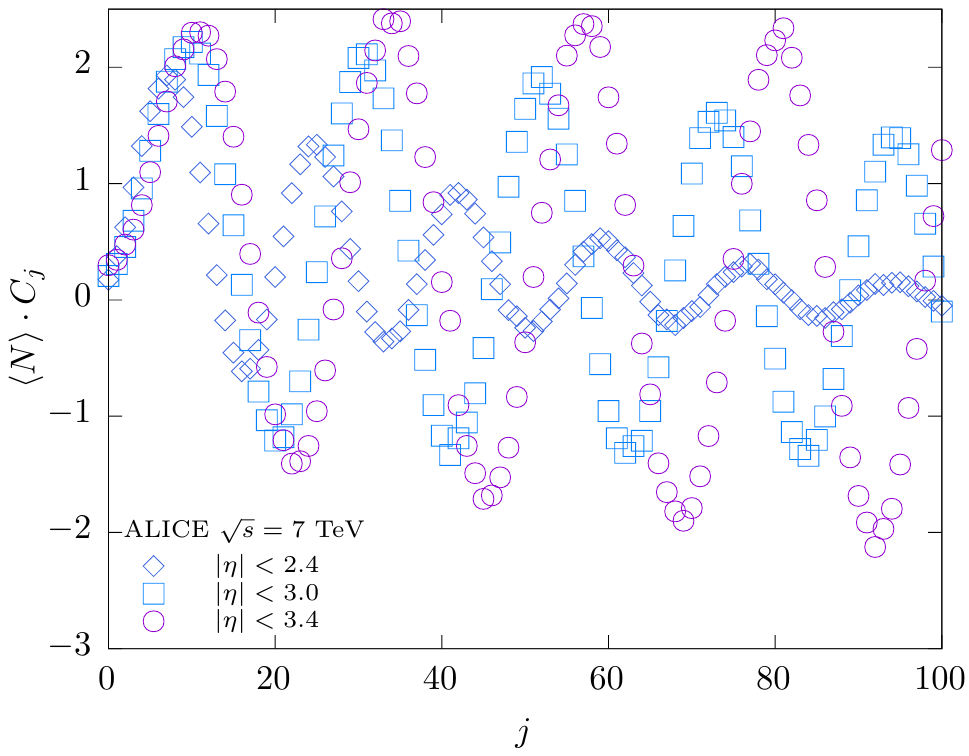}
\end{center}
\vspace{-5mm}
\caption{
Top panel: The plots of $C_j$ oscillations using $pp$ experimental data at $\sqrt{s}=7$ TeV derived from ALICE Collaboration over a pseudorapidity range up to $|\eta| < 2.4$ \cite{ALICE}. The magnitude and period is comparable to $C_j$ derived from the CMS Collaboration at the same energy and pseudorapidity. Bottom panel: $C_j$ plots from ALICE Collaboration \cite{ALICE8TeV.wide.eta} obtained for pseudorapidity up to $|\eta|<3.4$ plotted separately for clarity. Note the increase in oscillatory magnitude at $|\eta|<3.4$.}
\label{graph.pp.across.eta}
\end{figure}

\subsection{Dependence on pseudorapidity window}
\label{sect.eta.dependence}

\begin{figure}[t]
\begin{center}
\includegraphics[scale=0.75]{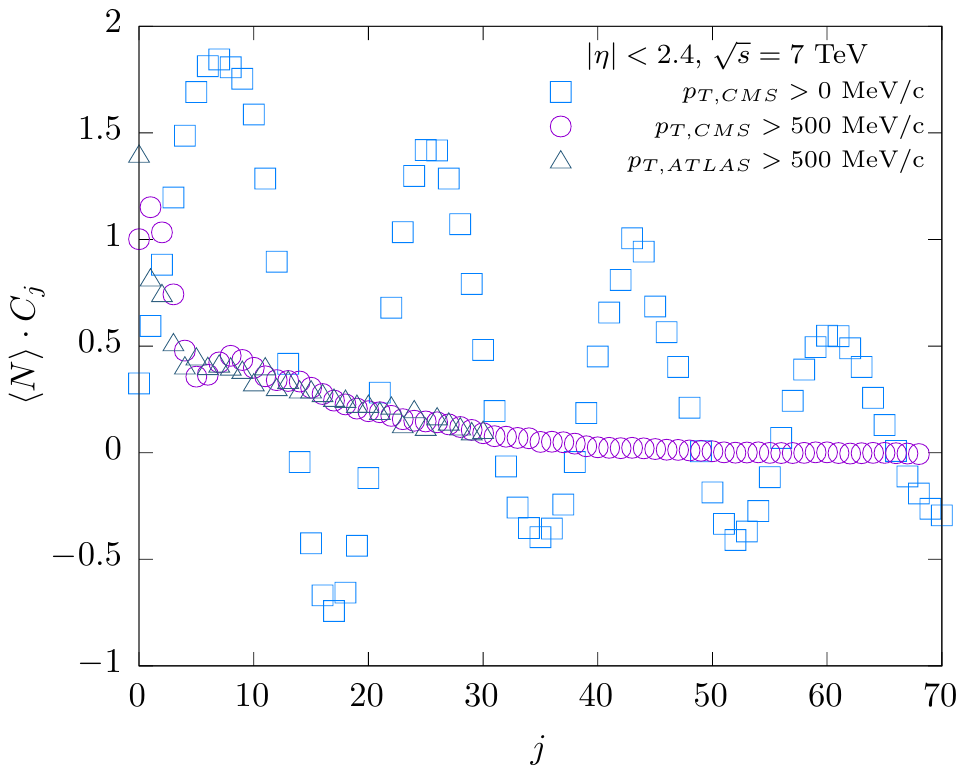}
\includegraphics[scale=0.85]{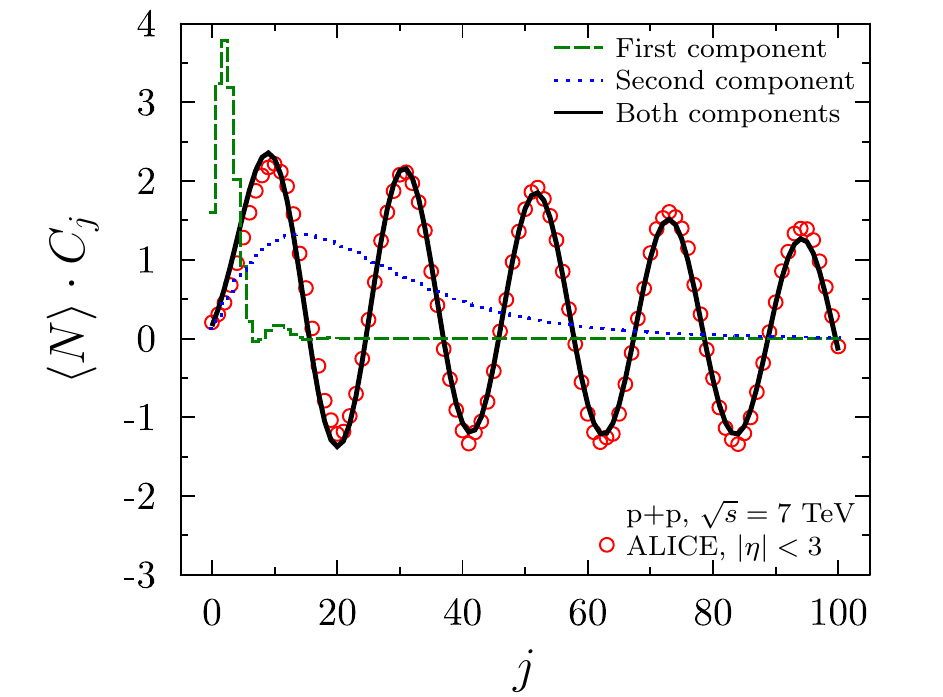}
\end{center}
\vspace{-5mm}
\caption{Top panel: The plot of $C_j$ vs $j$ with CMS data at $7$ TeV with $|\eta| < 2.4$ \cite{CMS7TeV} and ATLAS data \cite{ATLAS7TeV} with $\eta < 2.5$. CMS has extrapolated its data all the way to $p_T > 0$ MeV/c in the cited reference. This allows us to compare it with the data obtained experimentally with $p_T > 500$ MeV/c, also from CMS. The $C_j$ derived from ATLAS data tracks that of CMS closely. Bottom panel: Schematic view of modified combinants $C_j$ for separate components from the two component compound distribution given by Eqs. (\ref{2CBD}) and (\ref{2-CBD}) with parameters fitting experimental $P(N)$ shown in Fig. \ref{F1b}.}
\label{graph.pp.across.pt}
\end{figure}

The dependence of the extent of oscillations on the pseudorapidity window from which the experimental data was obtained is the most obvious. Intuitively, one would expect experimental data collected from a larger pseudorapidity phase space to be more representative of the collective behaviour of the underlying collision (e.g. $e^+e^-, pp$ or $p\bar{p}$) and the associated secondary particles.

Fig. \ref{graph.pp.across.eta} shows example of the observed differences between the oscillations in $C_j$ derived from different rapidity windows by ALICE Collaboration \cite{ALICE}.  The first observation is that oscillations, which are almost non-existent at small pseudorapidity window ($|\eta|<1.5$) are becoming very strong at the maximal pseudorapidity window ($|\eta| < 3.4$.

There is also a change in the period of oscillations (where present) with a change in pseudorapidity window. In general, the period decreases from around $18$ for $| \eta | < 2.4$, to approximately $11$ for $|\eta|<1.5$. The amplitude of oscillations for any smaller pseudorapidity window is too weak to discern the period. Nevertheless, the oscillations for the data from the ALICE Collaboration are relatively smooth within the pseudorapidity phase space.

 Data from the ALICE Collaboration had been taken over a larger pseudorapidity window, up to $\eta < 3.4$. This allows the investigation of behaviour of $C_j$ oscillations beyond the limited window $|\eta| \leq 2.4$ available in by CMS data (this is due to challenges surrounding the drastic drop in reconstruction efficiencies at $| \eta | > 2.4$ \cite{CMS7TeV}). The bottom panel of Fig. \ref{graph.pp.across.eta} has been plotted using ALICE data from $\eta < 2.4$ to $\eta < 3.4$ for better clarity. The trend of increasing period with larger pseudorapidity window continues beyond $|\eta|<2.4$. However, the rate of amplitude decay slows significantly between $|\eta|<2.4$ and $|\eta|<3.0$ and reverses at $|\eta|<3.4$. From this observation, it is inferred that the amplitude stops its decay and reversed somewhere between $3.0 < |\eta| < 3.4$.

\subsection{Dependence on $p_T$}
\label{pt.dependence}

In earlier study presented in \cite{3componentNBD} it was noted that the $C_j$ obtained from data obtained for $p_T > 100$ MeV/c cut by ATLAS \cite{ATLAS13TeV} exhibit minimal oscillation for $| \eta | < 2.5$, which are completely absent for data with $p_T > 500$ MeV/c cut. This observation suggests that the $p_T$ phase space plays a role in the extent of $C_j$ oscillations as well. For this subsection, we will consider $pp$ collision data obtained from the ATLAS and CMS collaborations across different $p_T$ cuts at $\sqrt{s} = 7$ TeV. The CMS collaboration performs an extrapolation down to $p_T = 0$ MeV/c for their MD data. This allows further exploration of the behaviour of the $C_j$ oscillations over the complete $p_T$ phase space. The resulting uncertainty due to the extrapolation is less than $1\%$, inclusive of systematic uncertainty \cite{CMS7TeV}.

Top panel of Fig \ref{graph.pp.across.pt} presents results for data at $\sqrt{s}=7$ TeV, from CMS at $|\eta| < 2.4$ and from ATLAS at $|\eta| < 2.5$. The small difference in the pseudorapidity window over which they are obtained is considered insignificant, as can be seen in the close tracking of the data points from CMS and ATLAS for $p_T > 500$ MeV/c. Note that the $C_j$ oscillations are the strongest at $p_T > 0$ MeV/c (from CMS) while having minimal oscillations at $p_T > 500$ MeV/c (both CMS and ATLAS). Due to the lack of availability of data points with consecutive integral $N$ from ATLAS at $p_T > 500$ MeV/c, the plot has to be truncated at $j=30$. Unfortunately, no $p_T$ data is available from earlier experiments. The dearth of such data prohibits further investigation into the effects on oscillations between various $p_T$ cuts in $p\bar{p}$ collisions. Nevertheless, even these limited results can be very helpful in understanding the message of $C_j$. They are very similar to what is presented in the bottom panel of Fig \ref{graph.pp.across.pt} which shows schematic view of modified combinants $C_j$ for separate components from the two component compound distribution given by Eqs. (\ref{2CBD}) and (\ref{2-CBD}) with parameters fitting experimental $P(N)$ shown in Fig. \ref{F1b}. This comparison seems to suggest that particles with large transverse momenta mainly come from the first component (with smaller mean multiplicity) in our two component compound distribution. In other words, top panel of Fig. \ref{graph.pp.across.pt} seems to show that reducing the $p_T$ phase space eliminates (at least to some extent) one of the components.

\subsection{Dependence on $\sqrt{s}$}
\label{energy.dependence}

The reason why data from $\sqrt{s}=7$ TeV has been extensively exploited in the earlier parts of this work is due to the fact that oscillatory behaviour are more apparent at higher collision energies. Hints of this potential dependence on collision energy can first be observed in Fig. \ref{F1c} between $p\bar{p}$ collisions at $\sqrt{s}=200$ GeV vs $\sqrt{s}=900$ GeV in similar pseudorapidity windows.

\begin{figure}[h]
\centering
\includegraphics[scale=0.75]{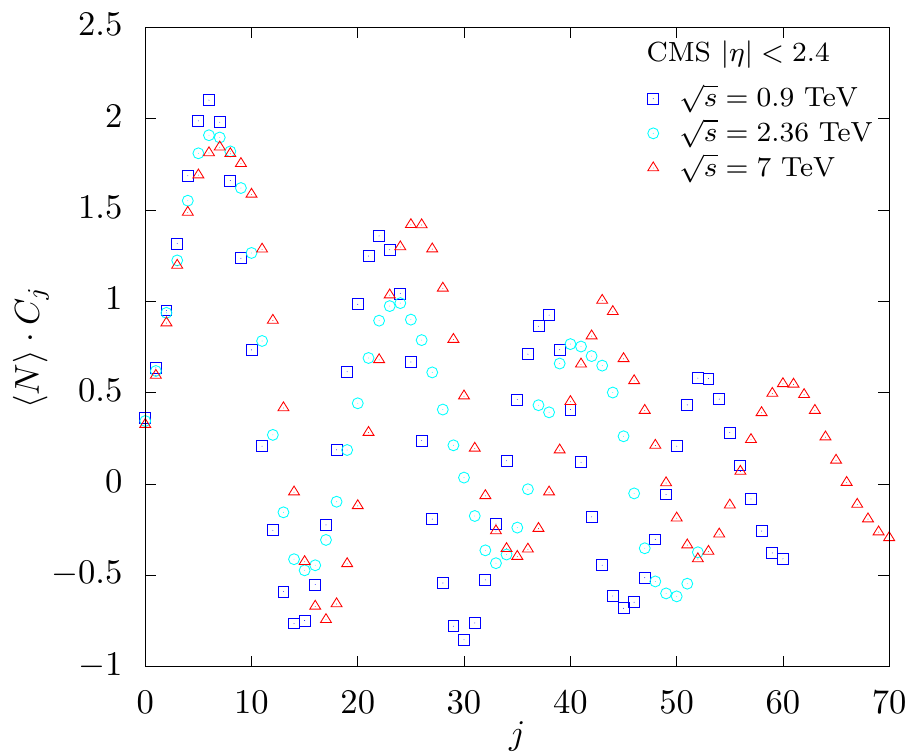}
\includegraphics[scale=0.75]{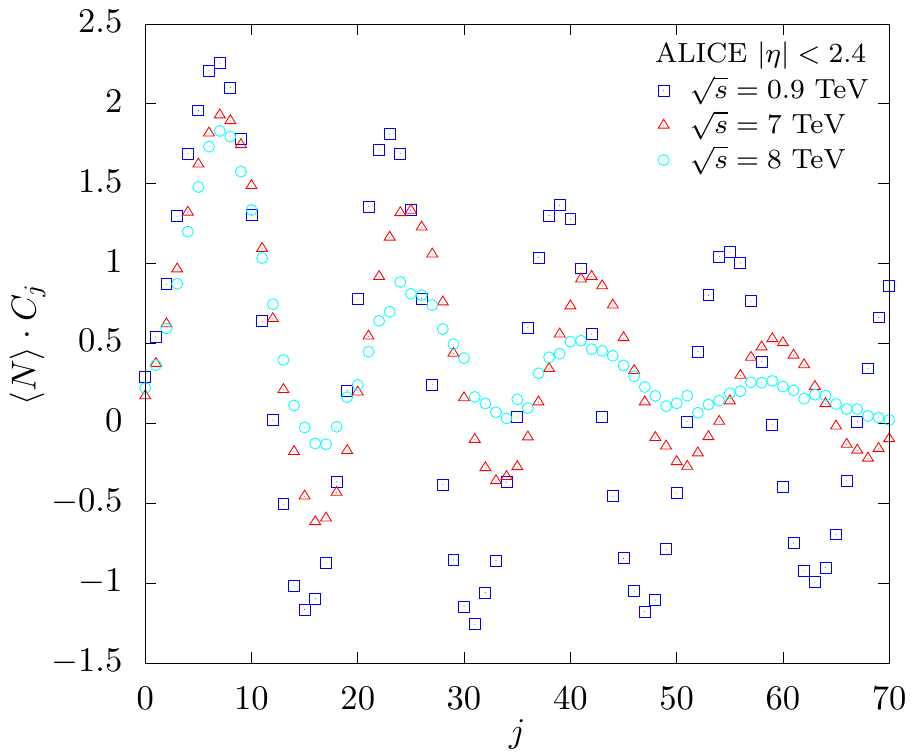}
\caption{Plots of $C_j$ vs $j$ across various centre-of-mass collision energies. Top panel: The plots of $C_j$ vs $j$ using data form CMS up to $\sqrt{s}=7$ TeV. It shows that the effect of an increase in centre-of-mass collision energies has minimal effect on the amplitude and the period of the resulting oscillations. Bottom panel: Plots of $C_j$ vs $j$ made using data from ALICE up to $\sqrt{s} = 8$ TeV. The amplitude seemed to undergo a much faster decay with an increase in collision energy, together with an increase of the oscillation period.}
    \label{graph.energy.dependence}
\end{figure}

The modified combinants derived form CMS \cite{CMS7TeV} and ALICE \cite{ALICE8TeV.wide.eta} across centre-of-mass energies are plotted in Fig. \ref{graph.energy.dependence} on the top and bottom panel respectively. The difference between the data sources is that ALICE provides data up to $\sqrt{s}= 8$ TeV while that from CMS is up to $\sqrt{s}=7$ TeV. To facilitate comparison, only data at $|\eta| < 2.4$ is used, on considerations that it shows the most distinct oscillatory behaviour without the amplitude blowing up. Note that CMS does not provide data obtained from wider pseudorapidity windows, which makes comparison difficult.

For the $C_j$ from CMS, there is no clear effect on the amplitude with increasing collision energies. The $C_j$ made with data from lower collision energy of $\sqrt{s} = 0.9 $ TeV appeared to have a slightly higher initial amplitude but decayed at similar rates to that from $\sqrt{s} = 7$ TeV. There is also an increase in the period of oscillation at higher energies. On the other hand, the graph derived from ALICE data seemed to show a more distinct difference in the amplitude between data from $\sqrt{s} = 0.9$ TeV and that from $\sqrt{s} = 7$ TeV with a slower rate of decay. The shorter period at lower energy is also observed here, and is consistent up to $\sqrt{s} = 8$ TeV.

However, there are some observed differences in the details between both plots in Fig. \ref{graph.energy.dependence}. Some examples include the higher amplitudes of the oscillating $C_j$ for ALICE than than CMS at the same $\sqrt{s}$, the location of the amplitudes with respect to $q$ etc. These discrepancies can be traced back to different $P(N)$ values obtained between the two experiments due to slightly different methods in which measured data is being treated between the two experiments. A more detailed comparison can be found in \cite{ALICE8TeV.wide.eta}. However, this difference should not mask the trend in $C_j$ oscillations with  increasing energies, which is the main point behind the plot.

\subsection{Summary and discussion of results}
\label{sect.discussion}

The pseudorapidity window within which the data has been obtained appears to have the most significant effect on the oscillatory period for the corresponding derived value of $C_j$. This feature can be clearly observed in the plot across various pseudorapidity windows from ALICE data in Fig \ref{graph.pp.across.eta}. There is  direct correspondence between the size of the pseudorapidity window to the oscillation period. While $C_j$ up to $|\eta| < 1.0$ barely exhibits any oscillations, the Top panel of Fig \ref{graph.pp.across.eta} shows an increase in period from 11 at $|\eta| < 1.5$ to 18 for $|\eta| < 2.4$. With reference to the bottom panel in Fig \ref{graph.pp.across.eta} for large pseudorapidity windows, we see that increasing the size of the window results in a corresponding increase in oscillatory period, from $18$ at $|\eta| < 2.4$ up to $23$ at $|\eta| < 3.4$. Data from UA5 paints a different story. The $C_j$ oscillates with period 2 at $\sqrt{s} = 900$ GeV at $|\eta| < 3.0$ and above. Coupled with the modified combinants derived from $e^+e^-$ \cite{e+e-}, this seems to suggest that $C_j$ from matter-antimatter ($p\bar{p}$ and $e^+e^-$) collisions oscillates more violently at comparatively lower energies than their $pp$ counterparts. This may be a feature useful in distinguishing between the two types of collision data.

The second effect of larger pseudorapidity window is on the amplitude of the oscillations of $C_j$. Referring to Fig \ref{graph.pp.across.eta}, the amplitude of oscillation increases from just below $1.5$ for data from $\eta < 1.5$, to around $1.8$ for $C_j$ derived form $\eta < 2.4$.

\begin{figure}[t]
\begin{center}
\includegraphics[scale=0.75]{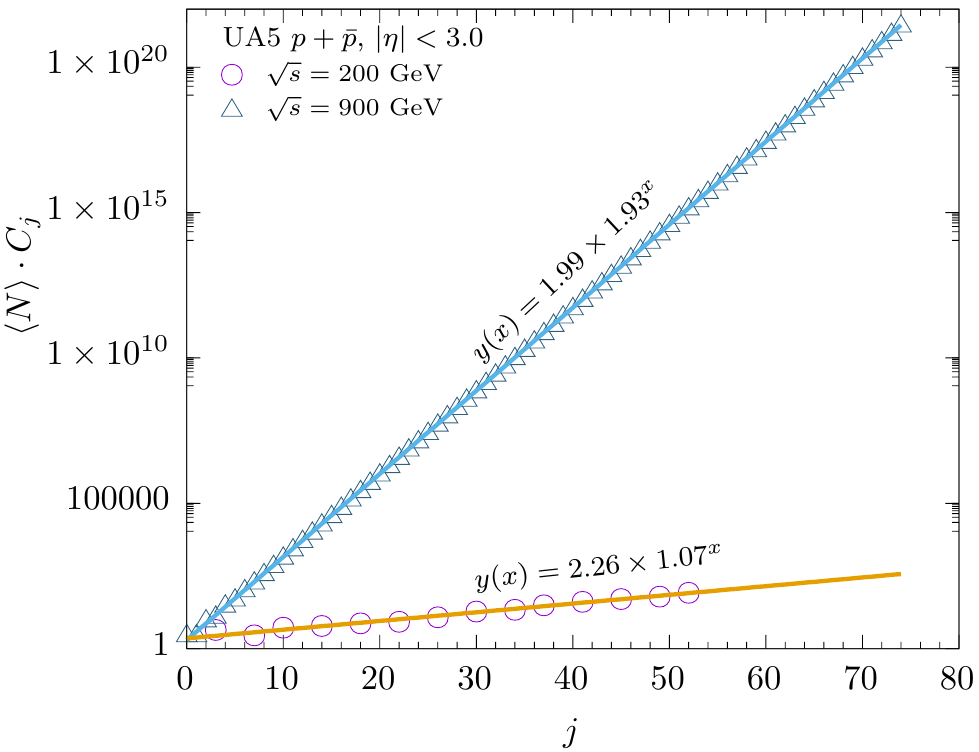}
\includegraphics[scale=0.75]{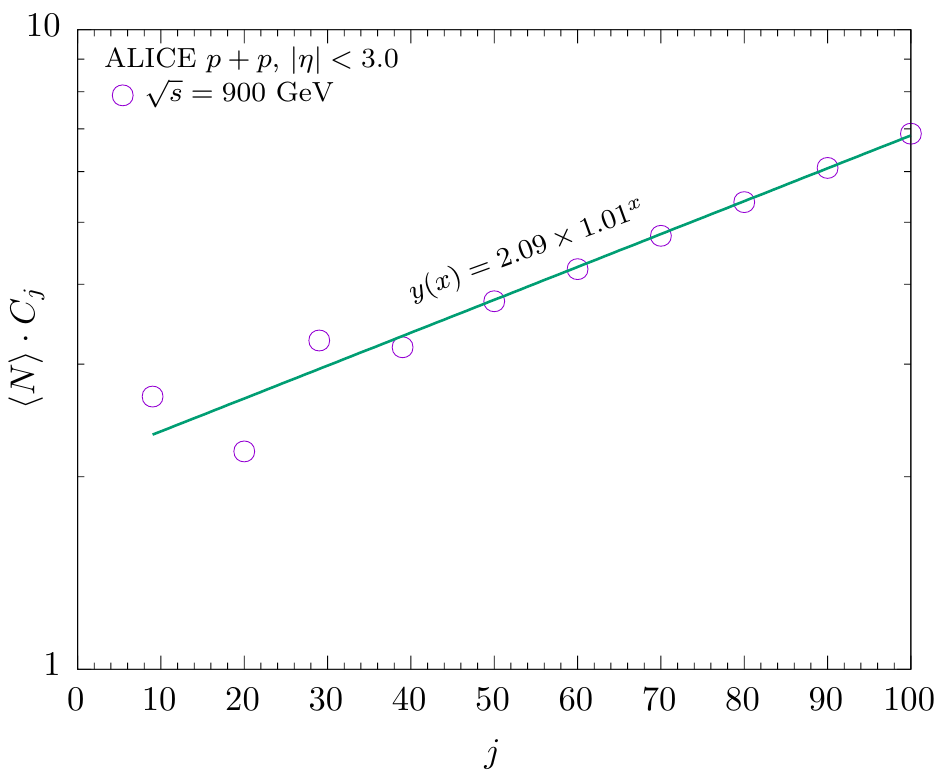}
\end{center}
\vspace{-5mm}
\caption{Top panel: The amplitude of the absolute values of $C_j$ are plotted against $j$ and fitted to a line $y(x) = 2.26\times1.07^x$ for $\sqrt{s} = 200$ GeV data, and to $y(x) = 1.99\times1.93^x$ for $\sqrt{s} = 900$ GeV data.
Bottom panel: The same for ALICE data at $\sqrt{s}= 900$ GeV and for $|\eta| < 3.0$. In this case the data points are fitted against a line $y(x) = 2.09\times1.01^x$. In both cases, the oscillation amplitude increases in a power-law fashion as function of $x=j$.}
\label{graph.ppbar.power.law}
\end{figure}

In Fig \ref{graph.ppbar.power.law} it is observed that both in UA5 and ALICE data the amplitudes of oscillations increase as a power-law from $\eta < 3.0$ onwards. The increase is more prominent for higher energies and for $p\bar{p}$ data from UA5. Note that when we use $G(z)$ as given by Eq. (\ref{GBDNBD}) then amplitude of oscillations is given by $[p/(1-p)]^j$. If the modified combinants were to be interpreted as weights of the various $P(N)$'s, as discussed in Section \ref{sect.mod.combinant}, the oscillations in the weights are more pronounced and periodic in a larger pseudorapidity phase space.

Another aspect which the oscillatory behaviour can be discussed is in terms of the $p_T$ phase space. In the top panel of Fig \ref{graph.pp.across.pt}, results from both CMS and ATLAS data shows an unambiguous relation between $p_T$ phase space and oscillatory extent of $C_j$. The extrapolation of the CMS data from $p_T > 0$ MeV/c for $\sqrt{s} \leq 7$ TeV allows us access to full $p_T$ phase space for LHC Run 1 energies. By comparing the derived $C_j$ from both CMS, ALICE and ATLAS, it is clear that like pseudorapidity, the larger the $p_T$ phase space, the larger the extent of oscillations. The comparison of these results with view of $C_j$ from separate components of distribution used to fit experimental $P(N)$ shown in Fig. \ref{F1b} which seems to suggest that particles with large transverse momenta mainly come from the first component is very instructive and suggest further investigations which, however, go beyond the goals of this work.

On the other hand, varying the collision energies does not produce such drastic changes in the extent of oscillations as compared to pseudorapidity and $p_T$ cuts. In Fig \ref{graph.energy.dependence}, we see that the effects of an increase in collision energy has minimal effects to the amplitude decay and the period of oscillatory behaviour. Both the amplitude and period of oscillations do not change significantly from $\sqrt{s}= 0.9 $ to $7 $ TeV for CMS, and up to $\sqrt{s} = 8 $ TeV for ALICE.

Note that usually the oscillatory behaviour of $C_j$ (as well as the lack of oscillations) is observed in the ideal cases, i.e., for $P(N)$ described by analytical formulas. Experimentally $C_j$'s are obtained from the measured multiplicity distributions, which are recorded with some acceptance in limited phase space and which contain both the systematic and statistical uncertainties. However, as was shown in \cite{e+e-}, experimental acceptance do not generate oscillations of $C_j$. For example, acceptance procedure applied to NBD gives again the NBD with the same $k$ but with the modified $p$, which is now equal to $p' = p\alpha/(1 - p +p\alpha)$, where $\alpha$ denotes the probability of the detection of a particle in the selected phase space. For a distribution described by NBD the acceptance does not alter the smooth decrease of $C_j$. In addition, the influence of statistical and systematic uncertainties on $C_j$ was discussed in details in earlier works \cite{MWW1, 3componentNBD}. It was found that at sufficiently-high statistics, modified combinants $C_j$ become relatively insensitive to statistical uncertainties, although the effects of systematic uncertainties of the measurements still remain.



However, it turns out that, notwithstanding this sensitivity, the oscillatory signal observed in the modified combinants derived from ATLAS, ALICE, CMS and UA5 data remains statistically significant. Therefore, the regularity and periodicity of the observed oscillations cannot be results of random fluctuations but instead, justify detailed and careful analysis of oscillations in modified combinants in the study of multiplicity distributions.

\section{Conclusions}
\label{sect.conclusion}

The utility of a phenomenological approach to analysis of multiplicity distributions stems from the lack of a comprehensive theoretical explanation transcending the hard and soft regimes of QCD. If enlarging the pseudorapidity phase space results in more distinct oscillatory behaviour, then the $C_j$ oscillations could find their origins in soft hadronic collisions.

This paper discusses dependence of $C_j$ oscillations on collision systems and the impact of varying pseudorapidity, $p_T$ cuts and collision energies on the oscillatory behaviour of $C_j$. It is clear that pseudorapidity has the greatest impact on the oscillatory behaviour among the experimental variables considered. The general trend inherent in the data shows increased oscillatory behaviour with an increase in the extent of phase space under considerations. Sampling within a larger extent of experimental phase space allows the collection of information from a larger domain. This in turn implies more representative data to be collected when the extent of phase space is large.

The way the $C_j$ oscillates between $pp$ and $p\bar{p}$ collisions is clearly different, in terms of the order of magnitude as well as the period. For $pp$ data from ALICE, the $C_j$ oscillates with a period of $20$. This is close to the earlier discussion in Section \ref{sect.eta.dependence} with $C_j$ oscillating at a period of $18$ at $\sqrt{s}=7$ TeV, $|\eta|<2.4$. In the case of $p\bar{p}$, $C_j$ oscillates with a period of $2$. Such a short period is reminiscent of our earlier work \cite{e+e-} exploring $C_j$ oscillations derived from  $e^+e^-$ collisions at $\sqrt{s} = 91$ GeV. Based on these two observations, it seemed that at sufficiently wide pseudorapidity window, $C_j$ from particle-antiparticle collisions at different energies oscillates with period $2$, while that from particle-particle collisions do not exhibit such regularity. Such power-law increase in amplitude may potentially be a characteristic of matter-antimatter collision, including that from $e^+e^-$.

Another distinguishing feature between $pp$ and $p\bar{p}$ collisions is the order of magnitude over which the oscillations take place. At $\sqrt{s}=900$ GeV, $C_j$ from $p\bar{p}$ goes up to a magnitude of $10^{20}$ while that for $pp$ stays below $10$. Should more data between the two types of collisions become available in the future, such figures can be tabulated to explore the dependence of the scaling coefficients on energy and pseudorapidity.

The relationship between $C_j$ and $F_q$ and $K_q$ moments as discussed in the \ref{appendixA} may offer some clues as to why $C_j$ derived from experimental MD data oscillates. The $H_q=K_q/F_q$ moments, with its roots in gluodynamics \cite{origins.H.moment,application.H.moment}, were conceived of and observed to undergo oscillations in earlier studies. On the other hand, $F_q$ has shown to be a valuable a tool in the study of intermittent behaviour \cite{origins.intermittency} in multiparticle production. Any attempts at a physical interpretation of $C_j$ can be considered in analogy to the relationship between $H_q$ and $F_q$. However, before that, the exact physical interpretation of $C_j$ still remains open and is subject for further investigation.

Finally, we will refer to the imposing question: what lesson can be learned from the behavior of modified combinants $C_j$ deduced from the measured multiplicity distributions $P(N)$ in what concerns the the dynamics of the multiparticle production process. First, it seems that the oscillations of the $C_j$ are closely related to
the need to use some specific form of multiplicity distribution (MD) in the description of these processes. It must be a compound distribution based on BD (CBD), which gives oscillations, with some other MD, which controls their period and amplitudes. In fact, as shown in \cite{MWW1} the successful use of simple sum of $3$ NBDs presented in \cite{3componentNBD} is possible only because such sum acts effectively as a kind of BD. In our investigations we were usually using MD which were either compound distributions of BD with NBD (either the sum of two such compound distributions to get perfect agreement with data)  or MD  for the sum of multiplicities from BD (or CBD) and NBD. In all cases BD is crucial to describe the oscillatory behaviour of modified combinants. This result, if taken seriously, imposes certain restrictions on the selection of the appropriate multi-particle production model. In \ref{appendixB} we present a summary of two potential candidates for such model, both based on some specific stochastic approach, one of which was used in this work. A broader discussion on this topic, in particular what other classes of models can meet the criteria required here, would require a separate work.

\begin{acknowledgements}
We are indebted to Edward Grinbaum-Sarkisyan  for fruitful discussions. This research  was supported in part by the National Science Center (NCN) under contracts 2016/23/B/ST2/00692 (MR) and 2016/22/M/ST2/00176 (GW). M. Ghaffar would like to thank NUS where part of this work is done for the hospitality. H.W. Ang would like to thank the NUS Research Scholarship for supporting this study. We would like to thank P. Agarwal and Z. Ong for reading the manuscript and for contributing to the insightful discussions.
\end{acknowledgements}

\appendix

\section{Relationship between $C_j$, $K_q$ and $F_q$ moments}
\label{appendixA}

A closely related quantities to modified combinants $C_j$ used to describe fluctuations in phenomenological studies \cite{book.soft.multihadron, book.fluctuations} is the set of factorial moments, $F_q$, and cumulant factorial moments $K_q$. Both can be defined by the multiplicity distributions, $P(N)$, and modified cumulants, $C_j$. The basic quantity to start with is the generating function
\begin{equation}
    G(z)=\sum_NP(N)z^N     \label{eqn.gen.fn}
\end{equation}
from which $P(N)$ emerges as
\begin{equation}
P(N) = \frac{1}{N!}\frac{d^N G(z)^{(N)}}{d z^N}\bigg|_{z=0} \label{PNfromG}
\end{equation}
and combinants $C^*_j$ are defined as
\begin{equation}
C^*_j = \frac{1}{j!}\frac{d^j \ln G(z)}{d z^j}\bigg|_{z=0}, \label{CstarfromG}
\end{equation}
note that according to Eq. (\ref{relation.to.combinant}) $\langle N\rangle C_j = (j+1)C^*_{j+1}$.
Similarly, the respective derivatives taken at $z=1$ define factorial moments, $F_q$,
\begin{equation}
F_q = \frac{d^q G(z)}{d z^q}\bigg|_{z=1} \label{FqfromG}
\end{equation}
and cumulant factorial moments, $K_q$,
\begin{equation}
K_q = \frac{d^q \ln G(z)}{d z^q}\bigg|_{z=1}. \label{KqfromG}
\end{equation}
Continuing this presentation, note that similarly as $P(N)$ defines $F_q$,
\begin{equation}
F_q = \sum_{N=q}^{\infty}\frac{N!}{(N-q)!}P(N) \label{FqfrmPn}
\end{equation}
the $C^*_j$ defines $K_q$,
\begin{equation}
K_q = \sum_{j=q}^{\infty}\frac{j!}{(j-q)!}C^*_j. \label{KqfromCj}
\end{equation}
Note that $K_q$, share the additive property of $C_j$. As an example, for a random variable made up of a sum of other random variables each described by a generating function $G_j(z)$, the generating function of the sum is given by $G(z)=\prod_jG_j(z)$. In this case, the value of $K_q$ of the sum is the sum of the $K_q$ values of the individual components, similar to how the modified combinants behave. While culmulants are suited to study the densely populated region of phase space, modified combinants are better suited for the sparsely populated regions. This can be seen from Eq. (\ref{rCj}), which only requires a finite sum of $P(N-j)$ terms in the calculation of $C_j$.

Recurrence relation given by Eq. (\ref{rCj}) follows naturally from definition of $C_j$. Using Leibniz's formula for the $j^{th}$ derivative of the quotient of two functions, $x=\frac{G'(z)}{G(z)}$,
\begin{equation}
    x^{(j)}=\frac{1}{G}\Bigg[ G^{(j+1)}-j!\sum_{k=1}^{j}\frac{G^{(j-k+1)}}{(j-k+1)!}\frac{x^{(k-1)}}{(k-1)!}\Bigg],
    \label{eqn.leibniz}
\end{equation}
where ${G'(z)}/{G(z)}={d\ln G(z)}/{dz}$. Comparing Eq. (\ref{KqfromG}) and Eq.(\ref{eqn.leibniz}), it is clear that $K_{q+1} = x^{(q)}|_{z=1}$. Using Eq. (\ref{eqn.leibniz}) for modified combinants defined by Eq. (\ref{Cj.in.Gz}), one arrives at
\begin{equation}
    \langle N \rangle C_j = (j+1)\Bigg[\frac{P(j+1)}{P(0)}\Bigg]-\langle N \rangle\sum_{i=0}^{j-1}C_i\Bigg[\frac{P(j-i)}{P(0)}\Bigg] ,                    \label{Eq(6)}
\end{equation}
which is just Eq. (\ref{rCj}) used before.

On a separate note, a variant of the unnormalized factorial moment $F_q$ has proved useful in the study of intermittent behaviours in high energy collisions \cite{origins.intermittency}. It has been shown that if intermittent behaviours do indeed persist in the detected multiplicity spectra, the multiparticle production mechanism takes the form of a cascading process \cite{hadronic.intermittency} via relations in the scaled factorial moments.

\section{The possible origin of observed oscillations of $C_j$}
\label{appendixB}
\setcounter{equation}{0}
In \cite{SBW} as a model for the particle production was considered the so called birth process with immigration. The production process proceeds via emission of particles from an incident colliding particle (by a kind of bremsstrahlung process) which can further produce another particles (via the birth process). This specific branching process is defined by the following evolution equation:
\begin{eqnarray}
\frac{\partial P(n;t)}{\partial t} &=& \lambda_0 [ - P(n;t) + P(n-1;t)] + \nonumber\\
&& + \lambda_2 [ - n P(n;t) + (n-1) P(n-1;t)],  \label{MNBD}
\end{eqnarray}
where $P(n; t)$ is the distribution of the number of particles at $t$ (the parameter describing the
evolution of a particle system from the initial state, $ t = 0$ to the final state corresponding to the maximum value $t= T$, with $T$ being some energy dependent parameter chosen to reproduce the energy dependence of the observed mean multiplicity), $\lambda_0$ is the immigration rate in an infinitesimal interval $(t,t+dt)$ and $\lambda_2$ is the production rate of the birth process in the interval $(t, t+dt)$.

In \cite{e+e-} we have used specific, QCD based, realization of such approach based on the stochastic branching equation (describing the total multiplicity distribution of partons inside a jet, \cite{GMD}),
\begin{eqnarray}
\frac{dP(n)}{dt} &=& - \left( A n + \tilde{A} m\right) P(n) + A(n-1)P(n-1) + \nonumber\\
                && + \tilde{A m}P(n-1) =\nonumber\\
                &=& \tilde{A}m[ - P(n) + P(n-1)] + \nonumber\\
                && + A [-n P(n) + (n-1)P(n-1)], \label{GMD-ZW}
\end{eqnarray}
where $t$ is now the QCD evolution parameter,
\begin{equation}
t = \frac{1}{2\pi b}\ln \left[ 1 + ab \ln \left( \frac{Q^2}{\mu^2}\right)\right], \label{QCD-t}
\end{equation}
with $Q$ being the initial parton invariant mass, $\mu$ a QCD mass scale (in GeV), $N_c = 3$ (number of colors), and $N_f = 4$ (number of flavors) and $2\pi b = \left( 11 N_c - 2 N_f\right)$. Now $P(n)$ is the probability distribution of $n$ gluons and $m$ quarks (to be fixed) at QCD evolution, with $A$ and $\tilde{A}$ referring  to the average probabilities of the branching process: $g \to gg$, and $q \to qg$ respectively. The parameter $\xi = m\tilde{A}/A$ is related to the initial number of quarks in average sense. Comparing Eqs. (\ref{MNBD}) and (\ref{GMD-ZW}) we can identify evolution parameters in both approaches:
\begin{equation}
\tilde{A} m = \lambda_0\qquad {\rm and}\qquad A = \lambda_2. \label{identify}
\end{equation}

In both approaches, defined by Eqs. (\ref{MNBD}) and (\ref{GMD-ZW}) one has to define initial condition. For
a set number of initial particles, $P(n;t=0) = \delta_{n,k'}$, one gets $G(z;t=0) = z^{k'}$ (this is the case of the GMD discussed in \cite{e+e-}).  For initial condition for $P(n;t=0)$ chosen in a form of binomial distribution, with two new parameters, $\alpha$ representing the production rate of additional particles (fireballs, clusters or a kind of "excited hadrons") present at $t = 0$, and $K$ denoting their maximal number, one gets boundary condition
\begin{equation}
G(z;t=0) = \sum_{n=0}^{\infty} P(n; t=0) z^n = [ 1 + \alpha (z-1)]^K. \label{Boundary}
\end{equation}
(used in \cite{SBW}), which leads to the following generating function:
\begin{equation}
G(z) = [ 1 - \kappa(z-1)]^{-(K+\xi)} \{ 1 - [ \kappa(1-\alpha) - 1](z-1)\}^K, \label{GF-ZW}
\end{equation}
where $\kappa = \exp\left(\lambda_2 T\right) - 1$ and $\xi = \lambda_0/\lambda_2$. Note that this is simply just a product of generating functions for the BD and NBD,
\begin{eqnarray}
G(z) &=& G_{BD}\left[ z; p'=1-\kappa(1-\alpha); K \right]\cdot\nonumber\\
 &&\cdot G_{NBD}\left[ z; p = \frac{\kappa}{1+\kappa};k=K+\xi \right], \label{productG}
\end{eqnarray}
and the respective modified combinants are given by Eq. (\ref{Cjbdnbd}).

In the case when initially some complex objects (firebals, cluster, jets and so on) are produced and subsequently each of them produces secondary particles, the corresponding multiplicity distribution is described by an appropriate compound distribution. If initial objects (sources) are produced according to a BD and subsequent production process is defined by Eqs. (\ref{MNBD}) or (\ref{GMD-ZW}) with initial condition $P(n;t=0) = \delta_{n,0}$ (i.e., multiplicity distribution from sources is given by the NBD), we have final compound distribution defined by the generating function
\begin{equation}
H(z) = G_{BD}\left[ G_{NBD} ( z; p=\kappa, k=\xi); p' =\alpha, K\right] \label{HGBDGNBD}
\end{equation}
as given by Eq. (\ref{2-CBD}). In this case combinants $C_j$ oscillate with period equal to $\sim 2m$, where $m=\kappa \xi/(1-\kappa)$ denote mean multiplicity from single source.

\end{document}